\documentclass[11pt]{article}
\usepackage{fullpage}
\usepackage{amsmath}
\usepackage{amssymb}
\usepackage{amsfonts}
\usepackage{epsfig}
\usepackage[all,poly]{xy}
\usepackage[dvips]{color}

\newcounter{algoctr}

\newif\ifnotesw\noteswtrue
   {\ifnotesw\marginpar[\hfill\(\top\)]{\(\top\)}\fi}%
   {\ifnotesw\marginpar[\hfill\(\bot\)]{\(\bot\)}\fi}

\newcommand{\mnote}[1]%
    {\ifnotesw\marginpar%
        [{\scriptsize\begin{minipage}[t]{\marginparwidth}
        \raggedleft#1%
                        \end{minipage}}]%
        {\scriptsize\begin{minipage}[t]{\marginparwidth}
        \raggedright#1%
                        \end{minipage}}%
    \fi}

\newcommand{\ignore}[1]{}

\newcommand{\etal}{{\it et al. }}

\newsavebox{\given}
\savebox{\given}[1em]{\rule[-1.5ex]{.2mm}{4ex}}

\newtheorem{theorem}{Theorem}
\newtheorem{corollary}[theorem]{Corollary}

\newtheorem{fact}[theorem]{Fact}

\newcommand{\blackslug}{\rule{7pt}{7pt}}

\newcommand{\iverson}[1]{\lbrack\!\lbrack #1 \rbrack\!\rbrack}
\newcommand{\qed}{\hfill{\setlength{\fboxsep}{0pt}
\framebox[7pt]{\rule{0pt}{7pt}}}}

\renewcommand{\notin}{\ifmmode \not\in \else $\not\in$ \fi}
\newlength{\thislabel}
\newcommand{\labsize}[1]{\settowidth{\thislabel}{#1}}

\newcommand{\prf}{\par\noindent{\sl Proof } \hspace{.01 in}}
\newcommand{\zo}{\{0,1\}}
\newcommand{\zz}{\mathbb{Z}_{2}}

\newcommand{\lip}{\langle}
\newcommand{\rip}{\rangle}

\newcommand{\bra}[1]{\langle #1 |}
\newcommand{\ket}[1]{| #1 \rangle}
\newcommand{\braket}[2]{\langle #1 | #2 \rangle}

\newcommand{\teta}{\tilde{\eta}}

\DeclareMathOperator{\supp}{supp}

\title{
Mixing of Quantum Walks on Generalized Hypercubes
} 
\author{
{Ana Best}\\
{McGill University}
\and
{Markus Kliegl}\\
{Swarthmore College}
\and
{Shawn Mead-Gluchacki}\\
{SUNY Potsdam}
\and
{Christino Tamon}\footnote{Contact author: tino@clarkson.edu}\\
{Clarkson University}
}
\date{\today}

\begin{document}
\bibliographystyle{plain}
\maketitle

\begin{abstract}
We study continuous-time quantum walks on graphs which generalize the hypercube.
The only known family of graphs whose quantum walk instantaneously mixes to uniform is the Hamming graphs
with small arities.  
We show that quantum uniform mixing on the hypercube is robust under the addition of perfect matchings 
but not much else. Our specific results include:
\begin{itemize}
\item The graph obtained by augmenting the hypercube with an additive matching 
	$x \mapsto x \oplus \eta$ is instantaneous uniform mixing whenever $|\eta|$ is even, 
	but with a slower mixing time. This strictly includes Moore-Russell's result on the hypercube \cite{mr02}.
\item The class of Hamming graphs $H(n,q)$ is not uniform mixing if and only if $q \ge 5$.
	This is a tight characterization of quantum uniform mixing on Hamming graphs; previously, 
	only the status of $H(n,q)$ with $q < 5$ was known.
\item The bunkbed graph $\mathcal{B}_{n}(A_{f})$ whose adjacency matrix is
	$I \otimes Q_{n} + X \otimes A_{f}$, where $A_{f}$ is a $\mathbb{Z}_{2}^{n}$-circulant
	matrix defined by a Boolean function $f$, is not uniform mixing if the Fourier transform
	of $f$ has support of size smaller than $2^{n-1}$.  
	This 
	explains why the hypercube is uniform mixing and why the join of two hypercubes is not.
\end{itemize}
Our work exploits the rich spectral structure of the generalized hypercubes and relies
heavily on Fourier analysis of group-circulants. 

\vspace{.025in}
\end{abstract}


\section{Introduction}

Quantum walk on graphs is an important area of study in quantum information and computation for
at least two reasons.
The first is algorithmic; as a natural generalization of classical random walks, quantum walks offer
an alternative paradigm to develop new quantum algorithms (see \cite{ccdfgs03,fgg07}). 
The second reason is physical; quantum walks provide a potentially simpler method for implementing 
quantum computers. Although both arguments are arguably still being debated and investigated, 
research into the intrinsic properties of quantum walk continues to be an important step towards 
these algorithmic and physical goals.
An excellent survey on quantum walks is given by Kendon \cite{k06}. 
In this paper, we focus on the continuous-time quantum walk on finite graphs and study its mixing properties.

In their breakthrough paper \cite{mr02}, Moore and Russell showed that the continuous-time quantum walk on 
the hypercube is instantaneous uniform mixing; more importantly, it mixes faster than the classical walk. 
Most well-known graphs, however, do not exhibit quantum 
uniform mixing, as later results showed. These include graphs such as the complete graphs, the Cayley graphs of the symmetric group, 
and most even-length cycles (see \cite{abtw03,gw03,aabehlt07}). 
In contrast, classical random walks are uniform mixing on most well-behaved graphs. 
This suggests that, unlike the classical case, uniform mixing is a rare phenomenon in quantum walks,
and that the hypercube is an anomaly.
We investigate if this quantum uniform mixing phenomenon still exists in some natural
generalizations of the hypercube.

First, we consider the standard hypercube augmented with perfect matchings defined by additive shifts 
$x \mapsto x \oplus \eta$, for $\eta \in \zz^{n}$. 
How destructive are these added matchings to uniform mixing on the hypercube?
Surprisingly, we found that the resulting hypercube $Q_{n}^{\eta}$ is {\em still} 
instantaneous uniform mixing whenever $|\eta|$ is even, and, more interestingly, that it has a {\em slower} 
mixing time than the standard hypercube. 
Since the augmented matchings lower the diameter of the hypercube, this is yet another example of a 
counter-intuitive phenomenon in quantum walks. 
This shows that quantum uniform mixing on the hypercube is robust under additive matchings.
For $|\eta|$ odd, we discover that the quantum walk is uniform mixing provided it starts 
in the superposition 
$\frac{1}{\sqrt{2}}(\ket{0_{n}} + \ket{\eta})$. 

Second, we study the class of Hamming graphs $H(q,n)$ which are $n$-dimensional $q$-ary hypercubes
(see Biggs \cite{biggs}). 
Prior to this work, it was known that $H(q,n)$ is quantum uniform mixing if 
$q \in \{2,3,4\}$ (see \cite{cfhrtw07}).
We close this gap by showing that $H(q,n)$ is {\em not} quantum uniform mixing if $q \ge 5$, for any $n$.
This gives a tight characterization of quantum uniform mixing on Hamming graphs.

Third, we generalize the hypercube using its recursive construction by Cartesian products. 
The $(n+1)$-dimensional hypercube $Q_{n+1}$ is built by combining two $n$-dimensional hypercubes 
$Q_{n}$ by connecting their corresponding vertices. More formally, 
$Q_{n+1} = Q_{n} \oplus K_{2}$, where $K_{2}$ is the complete graph on two vertices.
The adjacency matrix of this Cartesian product is given by $A_{Q_{n+1}} = I \otimes A_{Q_{n}} + X \otimes I$. 
So, the connection between the two copies of $Q_{n}$ is specified by the (second) identity matrix $I$. 
We explore the effect on quantum uniform mixing when this connection is varied.

To this end, we consider the bunkbed graph $\mathcal{B}_{n}(A_{f})$ whose adjacency matrix is
$I \otimes A_{Q_{n}} + X \otimes A_{f}$, where $A_{f}$ is a $\zz^{n}$-circulant defined by a Boolean
function $f$ over $\zz^{n}$ (see Diaconis \cite{diaconis}). 
Much like a standard circulant matrix, the function $f$ defines the first row of the matrix
$A_{f}$ and the group operation of $\zz^{n}$ determines the rest of the rows of $A_{f}$.
So, $A_{f}$ defines (and generalizes) the connection between the two hypercubes $Q_{n}$.
For example, the standard hypercube $Q_{n}=\mathcal{B}_{n-1}(I)$ has $f(x) = \delta_{x,0_{n}}$
and the hypercube with an additive matching $Q_{n}^{\eta}=\mathcal{B}_{n-1}(A_{f})$ has $f(x)=1$ whenever 
$x=0_{n}$ or $x=\teta$, where $\eta = 1 \cdot \teta$ without loss of generality.
Our main result is that $\mathcal{B}_{n}(A_{f})$ is not uniform mixing whenever the Fourier transform 
of $f$ has small support, that is, $|\supp(\hat{f})| < 2^{n-1}$. An immediate corollary shows that
the graph-theoretic join of two hypercubes $Q_{n} + Q_{n}$ is not uniform mixing. 
Unfortunately, the small Fourier support size is not a necessary condition since 
$\mathcal{B}_{n}(Q_{n})$ is not uniform mixing even though $|\supp(\hat{f})| \ge 2^{n-1}$.

Our work exploits the rich spectral structure of the generalized hypercubes and relies
heavily on Fourier analysis of group-circulants. A more complete treatment of the latter
may be found in Diaconis \cite{diaconis}.

We summarize the known status of quantum uniform mixing on graphs along with our contributions in 
Figure \ref{fig:results}.

\begin{figure}[b] \label{fig:results}
\caption{Instantaneous Uniform Mixing on Various Graphs.}
\vspace{.1in}
\begin{minipage}{7in}
\begin{tabular}{|l|c|l|}	\hline
Family of Graphs	&	Mixing	&	Reference	\\	\hline \hline
Hamming graph $H(n,q)$, ~$q \in \{2,3,4\}$	&	Yes	& 	Moore-Russell \cite{mr02}, Carlson \etal \cite{cfhrtw07}	\\
Complete multipartite graph 	&	No	& 	Ahmadi \etal \cite{abtw03} \\	
Symmetric group		&	No	&	Gerhardt-Watrous \cite{gw03} \\
Cycles 		&	No\footnote{This was proved for a subclass of even-length cycles}	&	Adamczak \etal \cite{aabehlt07} \\ \hline
Hamming graph $H(n,q)$, ~$q \ge 5$	&	No	&	this work \\
$\mathcal{B}_{n}(A_{f})$, $|\supp(f)| \in \{1,2\}$	
	&	Yes\footnote{For $|\supp(f)|=2$, we require $\supp(f)=\{0_{n},a\}$ with $|a|$ odd.}	&	Moore-Russell \cite{mr02}, this work \\
$\mathcal{B}_{n}(A_{f})$, $|\supp(\hat{f})| < 2^{n-1}$	&	No	& 	this work \\	\hline
\end{tabular}
\end{minipage}
\end{figure}


\section{Preliminaries}

For a logical statement $\mathcal{S}$, the Iversonian $\iverson{\mathcal{S}}$ 
is $1$ if $\mathcal{S}$ is true, and $0$ otherwise.
Let $\mathbb{Z}_{m}$ denote the additive group of integers $\{0,\ldots,m-1\}$ modulo $m$.
For $a,b \in \zz^{n}$, let $a \oplus b$ denote the bit-wise exclusive OR of $a$ and $b$,
let $a \cdot b = \sum_{k=1}^{n} a_{k}b_{k}\pmod{2}$ denote the inner product modulo $2$,
and let the Hamming weight $|a|$ be the number of ones in $a$.
We let $e_{j} \in \zz^{n}$ denote the unit vector that is $1$ in position $j$ and zero elsewhere.
We use $I$ and $J$ to denote the identity and all-one matrices, respectively; we use
$X$ to denote the Pauli-$\sigma_{X}$ matrix.

The graphs $G=(V,E)$ we study are finite, simple, undirected, and connected. 
The adjacency matrix $A_{G}$ of a graph $G$ is defined as $A_{G}[u,v] = \iverson{(u,v) \in E}$.
In most cases, we also require $G$ to be vertex-transitive, that is, 
for any $a,b \in V$, there is an automorphism $\pi \in Aut(G)$ with $\pi(a)=b$. 
The Cartesian product $G \oplus H$ of graphs $G$ and $H$ is a graph whose adjacency matrix is
$I \otimes A_{G} + A_{H} \otimes I$. Let $K_{n}$ denote the complete graph on $n$ vertices.
Then, the binary $n$-dimensional hypercube $Q_{n}$ may be defined recursively as 
$Q_{n} = Q_{n-1} \oplus K_{2}$, for $n \ge 2$, and $Q_{1} = K_{2}$.
For more background on algebraic graph theory, we refer the reader to Biggs \cite{biggs}.

Next, we describe group-theoretic circulant graphs and Fourier analysis on $\zz^{n}$ \cite{diaconis}.
Let $\mathcal{G}$ be a finite group of order $m$ and let $f: \mathcal{G} \rightarrow \mathbb{C}$ be a 
class function over $\mathcal{G}$ (that is, $f$ is constant on the conjugacy classes of $\mathcal{G}$). 
Then, the $m \times m$ matrix defined by $A^{\mathcal{G}}_{f}[s,t] = f(ts^{-1})$ is called
a $\mathcal{G}$-circulant matrix defined by $f$. Moreover, $A^{\mathcal{G}}_{f}$ defines a 
$\mathcal{G}$-circulant graph if $f$ is a $\zo$-valued function that satisfies 
$f(e)=0$, where $e$ is the identity element, and $f(a^{-1})=f(a)$, for all $a \in \mathcal{G}$.
Here, the correspondence with Cayley graphs is recovered by letting the generator set be $\{a : f(a)=1\}$.
In this paper, we focus on the Abelian group $\mathcal{G} = \zz^{n}$.

\begin{figure}[t]
\begin{center}
\begin{equation}
\begin{bmatrix}
a & b & c & d & e & f & g & h \\
b & a & d & c & f & e & h & g \\
c & d & a & b & g & h & e & f \\
d & c & b & a & h & g & f & e \\
e & f & g & h & a & b & c & d \\
f & e & h & g & b & a & d & c \\
g & h & e & f & c & d & a & b \\
h & g & f & e & d & c & b & a
\end{bmatrix}
\end{equation}
\caption{Example of a $\zz^{n}$-circulant matrix: $n=3$;
the first row completely determines the matrix through the group operation of $\zz^{n}$.
The hypercube requires $b=c=e=1$ and $a=d=f=g=h=0$.
}
\end{center}
\end{figure}

For some basic facts of Fourier analysis over $\mathbb{Z}_{2}^{n}$,
let $f,g: \zz^n \rightarrow \mathbb{C}$ be arbitrary functions. 
The inner product of $f,g$ is defined as $\lip f,g\rip = \sum_{x} f(x)g^{\star}(x)$. 
The group characters of $\zz^{n}$ are given by $\chi_{a}(x) = (-1)^{a \cdot x}$, for $a,x \in \zz^{n}$,
and they satisfy $\lip \chi_{a},\chi_{b}\rip = 2^n\iverson{a = b}$.
With this, we can define the Fourier transform of $f$ at $a$ as
\begin{equation}
\hat{f}(a) = \lip f,\chi_{a}\rip = \sum_{x} f(x)\chi_{a}(x)
\end{equation}
while the inverse Fourier transform is given by 
\begin{equation}
f(x) = 2^{-n}\sum_{a} \hat{f}(a)\chi_{a}(x).
\end{equation} 
The support of $f$ is $\supp(f) = \{x: \ f(x) \neq 0\}$; similarly, the support of 
$\hat{f}$ is $\supp(\hat{f}) = \{a: \ \hat{f} \neq 0\}$.
The Convolution Theorem states that
\begin{equation}
\widehat{fg}(a) = \frac{1}{2^n} \sum_{b} \hat{f}(b)\hat{g}(a \oplus b).
\end{equation}
If $P: \zz^{n} \rightarrow [0,1]$ is a probability distribution, then $\hat{P}(0_{n})=1$.
Moreover, $P$ is the uniform distribution if and only if $\hat{P}(a) = 0$ for all $a \neq 0_{n}$.

If $G=(V,E)$ is a graph with adjacency matrix $A$, let $\ket{\psi(t)} \in \mathbb{C}^{|V|}$ be
a time-dependent amplitude vector over $V$. 
Then, the continuous-time quantum walk on $G$ is defined using Schr\"{o}dinger's equation as
\begin{equation}
\ket{\psi(t)} = e^{-it A} \ket{\psi(0)},
\end{equation}
where $\ket{\psi(0)}$ is the initial amplitude vector (see \cite{fg98}). 
The {\em instantaneous} probability of vertex $v$ at time $t$ is 
$p_{v}(t) = |\braket{v}{\psi(t)}|^{2}$.
We say $G$ is {\em instantaneous uniform mixing} if there is a time $t^{\star}$ such that the
quantum walk on $G$ satisfies 
$|\braket{v}{\psi(t^{\star})}|^{2} = 1/|V|$, for all $v \in V$.

\subsection{Fourier Analysis of the Hypercube}

We briefly review the Fourier analysis for the hypercube $Q_{n}$ \cite{mr02}. 
Since the adjacency matrix $A$ of $Q_{n}$ is a $\zz^{n}$-circulant, its eigenvectors are
the characters $\ket{\chi_{a}}$ (expressed in ket notation) where 
$\braket{x}{\chi_{a}} = \chi_{a}(x) = (-1)^{a \cdot x}$, for $a,x \in \zz^{n}$. 
The first row of $A$ is defined by a Boolean function $f$ where $f(x)=\iverson{|x|=1}$.
Thus, the eigenvalues of $A$ are given by
$\lambda_{a} = \sum_{x} f(x)\chi_{a}(x) = \sum_{j=1}^{n} \chi_{a}(e_{j}) = n - 2|a|$.
This follows from the theory of group circulants (see Diaconis \cite{diaconis}), 
but can also be verified directly.
Since the hypercube is vertex-transitive, we may assume that the start vertex is $\ket{0}$ 
(which corresponds to vertex $0_{n}$); moreover, $\ket{0} = 2^{-n}\sum_{a} \ket{\chi_{a}}$.

The quantum walk on $Q_{n}$ starting at $\ket{0}$ is given by
$\ket{\psi(t)} = e^{-itA}\ket{0} = 2^{-n}\sum_{a} e^{-it\lambda_{a}}\ket{\chi_{a}}$.
Viewing time $t$ as being {\em fixed}, we view the amplitude vector $\ket{\psi(t)}$ as a function 
of $a \in \zz^{n}$ and redefine $\psi_{t}(a) = \braket{a}{\psi(t)}$. 
By Fourier inversion, we see that $\widehat{\psi_{t}}(a) = e^{-it\lambda_{a}}$.
Since $P_{t}(a) = |\braket{a}{\psi(t)}|^{2} = \psi^{\star}_{t}(a)\psi_{t}(a)$, 
using the Convolution Theorem, we have
$\widehat{P_{t}}(a) = 2^{-n}\sum_{b} \widehat{\psi_{t}}(b)\widehat{\psi^{\star}_{t}}(a \oplus b)$. 
Since $\widehat{\psi_{t}}^{\star} = \widehat{\psi^{\star}_{t}}$, we obtain
\begin{equation} \label{eqn:hat-p}
\widehat{P_{t}}(a) = \frac{1}{2^{n}}\sum_{b} \exp(-it(\lambda_{b} - \lambda_{a \oplus b}))
\end{equation} 
To show that $P_{t}$ is uniform, it suffices to show $\widehat{P_{t}}(a)=\iverson{a = 0_{n}}$;
which was proved by Moore and Russell \cite{mr02}.


\section{Hypercube with Additive Matchings}

In the standard hypercube $Q_{n}$, whose vertices are the elements of $\zz^{n}$, two vertices $a,b$ 
are adjacent if $a \oplus b = e_{j}$, for some $j$; that is, $a$ and $b$ differ in exactly one coordinate.
For $\eta \in \zz^{n}$, we define the $(n+1)$-regular graph $Q^{\eta}_{n}$ to be the graph obtained 
from $Q_{n}$ by adding the matching $(a, a \oplus \eta)$, for all $a \in \zz^{n}$.

\begin{figure}[h]
\begin{center}
\epsfig{file=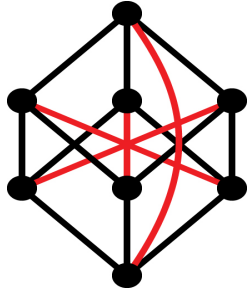, height=1.5in, width=1.25in}
\caption{Example of $Q_{n}^{\eta}$ hypercube: $n=3$ and $\eta = 111$.}
\label{figure:eta-cube}
\end{center}
\end{figure}

\begin{theorem} \label{thm:eta}
For $n \ge 2$,
a continuous-time quantum walk on $Q^{\eta}_{n}$ is instantaneous uniform mixing
if and only if $|\eta|$ is even. 
\end{theorem}
\prf
Let $A$ be the scaled adjacency matrix of $Q^{\eta}_{n}$, where $A_{a,b} = 1/(n+1)$
if $a \oplus b \in \{e_{1},\ldots,e_{n},\eta\}$, and $A_{a,b} = 0$ otherwise. 
Let $A_{0} = \{a \ | \ a \cdot \eta = 0\}$ and $A_{1} = \{a \ | \ a \cdot \eta = 1\}$.
Note $a \oplus b \in A_{0}$ if and only if $a,b \in A_{0}$ or $a,b \in A_{1}$.

\noindent By the group-circulant structure of $A$, its eigenvalues are given by
\begin{equation}
\lambda_{a} = \frac{1}{n+1}\left(n - 2|a| + (-1)^{a \cdot \eta}\right)
	= \left(1 - \frac{2k_{\eta}(a)}{n+1}\right)
\end{equation}
where
$k_{\eta}(a) = |a| + \iverson{a \in A_{1}}$.
\ignore{
k_{\eta}(a) = \left\{\begin{array}{ll}
		|a| & \mbox{ if $a \cdot \eta = 0$ } \\
		|a| + 1 & \mbox{ if $a \cdot \eta = 1$ }
		\end{array}\right.
}
We note that $k_{\eta}$ satisfies
\begin{equation} \label{eqn:delta-k}
k_{\eta}(a \oplus b) - k_{\eta}(b) 
\ignore{
	=
	\left\{
	\begin{array}{ll}
	|a \oplus b| - |b| 	& \mbox{ if $a \in A_{0}$ } \\
	|a \oplus b| - |b| + 1 	& \mbox{ if $a \in A_{1}$, $b \in A_{0}$ } \\
	|a \oplus b| - |b| - 1 	& \mbox{ if $a \in A_{1}$, $b \in A_{1}$ }
	\end{array}\right.
	=
	\left\{
	\begin{array}{ll}
	|a \oplus b| - |b| 				& \mbox{ if $a \in A_{0}$ } \\
	|a \oplus b| - |b| + (-1)^{b \cdot \eta} 	& \mbox{ if $a \in A_{1}$ }
	\end{array}\right.
}
	=
	|a \oplus b| - |b| + \iverson{a \in A_{1}}(-1)^{b \cdot \eta} 	
\end{equation}
Since the quantum walk on $Q_{n}^{\eta}$ is given by
$\ket{\psi(t)} = \frac{1}{2^n} \sum_{a} e^{-it\lambda_{a}}\ket{\chi_{a}}$,
we see that 
\begin{equation}
\widehat{\psi_{t}}(a) = \exp\left[-it\left(1-\frac{2k_{\eta}(a)}{n+1}\right)\right].
\end{equation}
A similar analysis to Equation \eqref{eqn:hat-p} yields
\begin{eqnarray}
\widehat{P_{t}}(a) 
	& = & \frac{1}{2^n}
		\sum_{b} \exp\left[-\frac{2it}{n+1}\left(k_{\eta}(a \oplus b) - k_{\eta}(b)\right)\right].
\end{eqnarray}
By Equation \eqref{eqn:delta-k}, we obtain
\begin{equation}
\widehat{P_{t}}(a) =
	\left\{\begin{array}{ll}
	1	&	\mbox{ if $a = 0_{n}$ } \\
\ignore{
	2^{-n}\sum_{b} \exp\left[-\frac{2it}{n+1}(|a \oplus b| - |b|) \right]
		&	\mbox{ if $a \in A_{0}$ } \\
	2^{-n}\sum_{b} \exp\left[-\frac{2it}{n+1}(|a \oplus b| - |b| + (-1)^{b \cdot \eta}) \right]
		&	\mbox{ if $a \in A_{1}$ }
}
	2^{-n}\sum_{b} \exp\left[-\frac{2it}{n+1}(|a \oplus b| - |b| 
			+ \iverson{a \in A_{1}}(-1)^{b \cdot \eta}) \right]
		& 	\mbox{ otherwise }
	\end{array}\right.
\end{equation}
For $a \in A_0$, analysis similar to that in \cite{mr02} shows that $\widehat{P_{t}}(a)$ is periodic in $t$
with period $(n+1)\pi$ and that, up to periodicity, $\widehat{P_{t}}(a) = 0$ only 
for $t_1^{\star} = (n+1)\frac{\pi}{4}$ and $t_2^{\star} = (n+1)\frac{3\pi}{4}$. 
This shows that the only possible times at which $Q_n^\eta$ could 
possibly be uniform mixing are $t_1^{\star}$ and $t_2^{\star}$. 
We will show for $t = t_1^{\star}$ that we indeed have 
$\widehat{P_{t}}(a) = 0$ for all $a \in A_1$ if and only if $|\eta|$ is even. 
The proof for $t = t_2^{\star}$ works in exactly the same way.

So assume that $a \in A_{1}$. Define
\begin{equation}
\rho_{a}(b) = |a \oplus b| - |b| + (-1)^{b \cdot \eta}.
\end{equation}
Observe that $\rho_{a}(a \oplus b) = -\rho_{a}(b)$.
Using this symmetry,
\begin{eqnarray}
\widehat{P_{t}}(a) 
	& = & \frac{1}{2^{n}}\sum_{b} 
		\frac{1}{2}\left[\exp\left(-\frac{2it}{n+1}\rho_{a}(b))\right)
		+ \exp\left(-\frac{2it}{n+1}\rho_{a}(a \oplus b)\right)\right] \\
	& = & \frac{1}{2^{n}}\sum_{b} \cos\left( \frac{2t}{n+1}\rho_{a}(b) \right)
\end{eqnarray}
At $t = t_1^{\star}$, we have
\begin{equation}
\widehat{P_{t_1^{\star}}}(a) = \sum_{b} \cos\left( \frac{\pi}{2} \rho_{a}(b) \right)
\end{equation}
Let $m$ be the number of overlaps of $1$'s between $a$ and $b$. Then
\begin{equation}
|a \oplus b| - |b| = |a| - 2m.
\end{equation}
Note that $m$ is even if $a \cdot b = 0$, and $m$ is odd if $a \cdot b = 1$.
Using these facts, we see that
\begin{eqnarray}
\sum_{b} \cos\left(\frac{\pi}{2} p_{a}(b)\right)
	& = & \sum_{b} \cos\left(\frac{\pi}{2}(|a|-2m+(-1)^{b \cdot \eta})\right) \\
	& = & \sum_{b} \cos\left(\frac{\pi}{2}(|a|+(-1)^{b \cdot \eta})\right)\cos(m\pi) \\
	& = & \sum_{b} (-1)^{a.b}\cos\left(\frac{\pi}{2}(|a|+(-1)^{b \cdot \eta})\right)
\end{eqnarray}
If $|a|$ is even, the last expression is $0$ and we are done.
Therefore assume for the rest of the argument that $|a|$ is odd. Then,
\begin{eqnarray}
\sum_{b} \cos\left(\frac{\pi}{2} p_{a}(b)\right)
	& = & \sum_{b} (-1)^{a.b}\cos\left(\pi\frac{(|a|+1)}{2} + \pi\frac{((-1)^{b \cdot \eta}-1)}{2}\right) \\
	& = & \sum_{b} (-1)^{a.b} (-1)^{\frac{|a|+1}{2}} (-1)^{b \cdot \eta} \\
	& = & (-1)^{\frac{|a|+1}{2}} \sum_{b} \chi_{b}(a \oplus \eta) \\
	& = & (-1)^{\frac{|a|+1}{2}} \iverson{a = \eta} \ 2^{n} \label{eq:phat}
\end{eqnarray}
Finally, notice that, since we are considering $a \in A_1$ such that $|a|$ is odd only, the case $a = \eta$ will occur if and only if $\eta \in A_{1}$ and $|\eta|$ is odd. Since the condition $\eta \in A_{1}$ is equivalent to $|\eta|$ being odd, this proves the claim.
\qed

\paragraph{Remark 1}
This includes the Moore-Russell result in the following sense. For $\eta = 0_n$, the Theorem shows that $Q_n^{\eta}$ is uniform mixing. If $\eta = 0_n$, however, all we are doing is adding self-loops to the standard hypercube, which just amount to scaling time by the factor $n/(n+1)$. Hence this shows that the standard hypercube is uniform mixing.

\paragraph{Remark 2}
The case when $|\eta|$ is odd is interesting as well. Namely, it is near uniform mixing in the following sense.
If $|\eta|$ is odd, taking the inverse Fourier transform of $\widehat{P_{t}}$ at time $t^\star_1$ or $t^\star_2$ yields
\begin{equation}
P_{t}(a) = \left\{
	\begin{array}{ll}
	\iverson{a \in A_0} ~ 2^{-n+1} & \mbox{ if $|\eta| \equiv 3 \pmod{4}$ } \\
	\iverson{a \in A_1} ~ 2^{-n+1} & \mbox{ if $|\eta| \equiv 1 \pmod{4}$ }
	\end{array}
	\right.
\end{equation}
That is, the probability distribution at these times is uniform on half of the vertices and zero on the other half.
Moreover, it can be shown that if the quantum walk is started in the superposition
$\ket{\psi(0)} = \frac{1}{\sqrt{2}}\left(\ket{0} + \ket{\eta}\right)$,
then $P_{t}$ is uniform at times $t_1^\star$ and $t_2^\star$, just as in the case when $|\eta|$ is even.


\section{Hamming Graphs}

As defined in Biggs \cite{biggs}, the {\em Hamming} graph $H(n,q)$ has vertex set $V = \{1,\ldots,q\}^{n}$ , and two vertices are adjacent if they differ in exactly one coordinate. As such, Hamming graphs are a 
natural generalization of the hypercubes $Q_{n}$ (which are simply $H(n,2)$). 

\begin{figure}[h]
\vspace{.1in}
\begin{center}
\epsfig{file=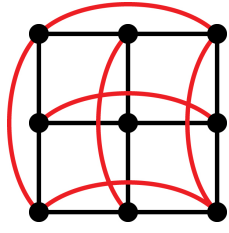, height=1.15in, width=1.15in}
\caption{Example of a Hamming graph: $H(2,3)$.}
\label{figure:hamming-graph}
\end{center}
\end{figure}

We prove that most of these Hamming graphs are not instantaneous uniform mixing. 
Before we prove this result, however, we need a preliminary result about quantum walks on the 
Cartesian product of graphs, which in turn relies on the following Fact.

\begin{fact}\label{fact:cartesian}
A quantum walk on $G \oplus H$ starting on vertex $(g, h)$ satisfies
\begin{equation}
\ket{\psi_{G \oplus H}(t)} = \ket{\psi_G(t)}\otimes \ket{\psi_H(t)},
\end{equation}
where $\psi_G$ and $\psi_H$ are quantum walks on $G$ and $H$ starting on vertices $g$ and $h$, respectively.
\end{fact}
\prf
The adjacency matrix of $G \oplus H$ is given by $I \otimes H + G \otimes I$. 
Since $I \otimes H$ and $G \otimes I$ commute, we have
\begin{eqnarray}
\ket{\psi_{G \oplus H}(t)} 
&=& e^{-it(I \otimes H + G \otimes I)} \ket{(g,h)}_{G\oplus H} \\
&=& e^{-it(I \otimes H)} e^{-it(G \otimes I)} (\ket{g}_G \otimes \ket{h}_H) \\
&=& (I \otimes e^{-itH})(e^{-itG} \otimes I)(\ket{g}_G \otimes \ket{h}_H) \\
&=& (e^{-itG} \ket{g}_G) \otimes (e^{-itH} \ket{h}_H) \\
&=& \ket{\psi_G(t)} \otimes \ket{\psi_H(t)}.
\end{eqnarray}
\qed\\

Here, it is convenient to look at the quantum walk on unnormalized adjacency matrices. 
This is permissible, since normalizing merely affects the time scaling. 
If $k_G$ and $k_H$ are the regularities of $G$ and $H$, respectively, Fact~\ref{fact:cartesian} 
may be written as
\begin{equation}
	\ket{\psi_{G\oplus H}\left((k_G+k_H)t\right)} 
	= \ket{\psi_G\left(k_G t\right)} \otimes \ket{\psi_H\left(k_H t\right)} 
\end{equation}
in the normalized case. Similar adjustments can be made throughout the rest of this paper.

In quantum mechanics, the natural way to combine two systems is through the tensor product. 
The previous Fact suggests that the Cartesian graph product serves a similar role for quantum walks.

The following Corollary is the key ingredient to proving our result about Hamming graphs. 
One direction of the Corollary was already proved in \cite{cfhrtw07}.

\begin{corollary}\label{cor:cpim}
A quantum walk on $G \oplus H$ starting on vertex $(g_0, h_0)$ is instantaneous uniform mixing at time 
$t^\star$ if and only if quantum walks on $G$ and $H$ starting on vertices $g_0$ and $h_0$, respectively, 
are instantaneous uniform mixing at time $t^\star$.
\end{corollary}
\prf
Fact \ref{fact:cartesian} shows that the amplitude at vertex $(g, h)$ is given by
\begin{align}
 (\bra{g}_G \otimes \bra{h}_H)(\ket{\psi_G(t)} \otimes \ket{\psi_H(t)})
 &= \braket{g}{\psi_G(t)} \braket{h}{\psi_H(t)}.
\end{align}
The probability at vertex $(g, h)$ is therefore simply
\begin{align}
P^{(g,h)}_{G \oplus H}(t) &= P^g_G(t)P^h_H(t).
\end{align}
If $P_G$ and $P_H$ are uniform at time $t^\star$, then clearly so is $P_{G \oplus H}$. If one of $P_G$ or $P_H$ is not uniform at time $t^\star$, then neither is $P_{G \oplus H}$. For suppose, without loss of generality, that $P^{g_1}_G \not= P^{g_2}_G$. Then for any nonzero $P_H^h$, we have
\begin{align}
 P^{(g_1, h)}_{G \oplus H} &= P^{g_1}_G P^h_H \not = P^{g_2}_G P^h_H = P^{(g_2, h)}_{G \oplus H},
\end{align}
showing $P_{G\oplus H}$ is indeed not uniform.
\qed\\

With the help of Corollary \ref{cor:cpim}, we may now easily prove our main Theorem of this section.
\begin{theorem}
For all $n \ge 1$,
a continuous-time quantum walk on the Hamming graph $H(n,q)$ is not instantaneous uniform mixing,
unless $q \le 4$.
\end{theorem}
\prf
It is known that $K_q$ is not uniform mixing unless $q \leq 4$ \cite{abtw03}. 
The proof therefore follows immediately from noting that
\begin{align}
H(n, q) &= \underbrace{K_q \oplus \cdots \oplus K_q}_{\text{$n$ times}},
\end{align}
and repeatedly applying Corollary~\ref{cor:cpim}.
\qed


\section{Bunkbed Variants}

In this section, we shall view the hypercube $Q_{n+1}$ as a bunkbed graph with adjacency matrix 
$I \otimes Q_{n} + X \otimes I$. 
To this end, we consider a bunkbed operator $\mathcal{B}_{n}(A)$
defined by
\begin{equation}
\mathcal{B}_{n}(A) = I \otimes Q_{n} + X \otimes A =
\begin{bmatrix}
Q_{n} & A \\
A & Q_{n}
\end{bmatrix} 
\end{equation}
where $A$ specifies the {\em connection} between the two copies of $Q_{n}$. 
We investigate the effect of this connection matrix $A$ on the quantum uniform mixing of $\mathcal{B}_{n}(A)$.

Let the connection graph $A$ be a $\zz^{n}$-circulant defined by a Boolean function $f: \zz^{n} \rightarrow \zo$. 
The eigenvalues of $A$ are given by $\lambda(a) = \sum_{x} f(x)\chi_{a}(x) = \hat{f}(a)$, for $a \in \zz^{n}$. 
For $a \in \zz^{n+1}$, we write $a = a_1 \cdot \tilde{a}$, where $a_1 \in \zz$ is the first bit of $a$ 
and $\tilde{a} \in \zz^{n}$ consists of the remaining bits of $a$.
The eigenvalue $\lambda_{a}$ of $\mathcal{B}_{n}(A)$ is given by
\begin{equation}
\lambda(a) 
	= \lambda_{Q_{n}}(\tilde{a}) + (-1)^{a_{1}}\hat{f}(\tilde{a}),
\end{equation}
In what follows, let $\Delta = \lambda(a \oplus b) - \lambda(b)$ and
$\tilde{\Delta} = \lambda(\tilde{a} \oplus \tilde{b}) - \lambda(\tilde{b})$.
Using similar analysis to Equation \eqref{eqn:hat-p}, we obtain
\begin{equation} \label{eqn:bunkbed}
\widehat{P_{t}}(a) 
	= \frac{1}{2^{n+1}} \sum_{b} \exp\left[ -it\left(\tilde{\Delta} 
	+ (-1)^{b_{1}}\left\{(-1)^{a_{1}}\hat{f}(\tilde{a} \oplus \tilde{b}) - \hat{f}(\tilde{b})
	\right\} \right) \right]
\end{equation}
First, we show a simple proof for the hypercube using this bunkbed framework.

\begin{fact} (Moore-Russell \cite{mr02})
For $n \ge 1$, $Q_{n+1} = \mathcal{B}_{n}(I)$ is uniform mixing.
\end{fact}
\prf
Since $A = I$, we have $f(x) = \iverson{x = 0_{n}}$; thus, $\hat{f}(a) = 1$, for all $a$. 
Using this in Equation \eqref{eqn:bunkbed}, we get
\begin{equation}
\widehat{P_{t}}(a) = \frac{1}{2^{n+1}}\sum_{b}\exp\left[-it\left(\tilde{\Delta} + (-1)^{b_{1}}((-1)^{a_{1}}-1)
	\right)\right].
\end{equation}
Therefore,
\begin{eqnarray}
\widehat{P_{t}}(0 \cdot \tilde{a}) 
	& = & \frac{1}{2^{n}}\sum_{\tilde{b}} \exp(-it\tilde{\Delta}) = \hat{P}_{t}(\tilde{a}) \\
\widehat{P_{t}}(1 \cdot \tilde{a}) 
	& = & \frac{1}{2^{n+1}}\sum_{b} \exp\left[-it(\tilde{\Delta} - 2(-1)^{b_{1}})\right] 
\ignore{
	& = & \frac{1}{2^{n+1}}\sum_{b_{1}=0} \exp\left[-it(\tilde{\Delta} - 2)\right] 
		+ \frac{1}{2^{n+1}}\sum_{b_{1}=1} \exp\left[-it(\tilde{\Delta} + 2)\right] \\
	& = & \frac{1}{2^{n}}\sum_{b} e^{-it\tilde{\Delta}}\cos(2t) \\
}
	= \cos(2t)\hat{P}_{t}(\tilde{a}).
\end{eqnarray}
By induction, this shows that $\widehat{P_{t}}(a) = \cos(2t)^{|a|}$. Thus, at $t^{\star} \equiv \pi/4$, 
we have $\widehat{P_{t^{\star}}}(a) = \iverson{a = 0_{n}}$, which proves that $P_{t^{\star}}$ is the
uniform distribution.
\qed\\

\par\noindent 
In the bunkbed framework, $Q_{n}^{\eta}$ is given by $\mathcal{B}_{n}(A_{f})$, 
where $\supp(f) = \{0_{n},\teta\}$; 
here, we assume without loss of generality that $\eta = 1 \cdot \teta$.
Next, we show that the bunkbed framework is a natural setting for showing limits on 
uniform mixing.

\begin{theorem} \label{thm:connection}
For $n \ge 1$, let $G = \mathcal{B}_{n}(A_{f})$ be a bunkbed hypercube, where $A_{f}$ is a $\zz^{n}$-circulant
defined by a Boolean function $f: \zz^{n} \rightarrow \zo$.
Then, $\mathcal{B}_{n}(A_f)$ is not uniform mixing whenever $|\supp(\hat{f})| < 2^{n-1}$.
\end{theorem}
\prf
\ignore{
Note that
\begin{equation}
\lambda_{A_f}(\tilde{a}) = \sum_{x} f(x)\chi_{\tilde{a}}(x) = \hat{f}(\tilde{a}).
\end{equation}
Since the adjacency matrix of $G$ is given by $I \otimes Q_{n} + X \otimes A_f$, the eigenvalues
of $G$ are given by:
\begin{equation}
\lambda(a_1 \cdot \tilde{a}) = \lambda_{Q_n}(\tilde{a}) + (-1)^{a_1}\hat{f}(\tilde{a}).
\end{equation}
}
Note that
\begin{eqnarray}
\lambda(a \oplus b) - \lambda(b)
        & = &
        \left[\lambda_{Q_n}(\tilde{a} \oplus \tilde{b}) - \lambda_{Q_n}(\tilde{b})\right] +
        (-1)^{b_1}\left[(-1)^{a_1}\hat{f}(\tilde{a} \oplus \tilde{b}) - \hat{f}(\tilde{b})\right] \\
        & = &
        \left[\lambda_{Q_n}(\tilde{a} \oplus \tilde{b}) - \lambda_{Q_n}(\tilde{b})\right] -
        (-1)^{b_1}\left[\hat{f}(\tilde{a} \oplus \tilde{b}) + \hat{f}(\tilde{b})\right], 
	\ \ \ \mbox{ if $a_1 = 1$ }
\end{eqnarray}
Thus, at $a = 1 \cdot 0_{n}$, we obtain
\begin{eqnarray}
\widehat{P_{t}}(1 \cdot 0_{n}) 
        & = & \frac{1}{2^{n+1}}\sum_{b} \exp\left[2it (-1)^{b_1} \hat{f}(\tilde{b})\right] 
        	= \frac{1}{2^{n}}\sum_{\tilde{b}} \cos\left(2\hat{f}(\tilde{b})t\right) \\
	& = & \frac{1}{2^{n}}\left[2^{n} - \left(|\supp(\hat{f})| - 
		\sum_{\tilde{b} \in \supp(\hat{f})} \cos(2\hat{f}(\tilde{b})t)  \right)\right].
\ignore{
        & = & \frac{1}{2^{n}}\left(|S|\cos(2Bt) + 2^{n} - |S| \right) \\
        & = & \frac{1}{2^{n}}\left[2^{n} - |S|(1 - \cos(2Bt)) \right] \ \neq \ 0,
}
\end{eqnarray}
When $|\supp(\hat{f})| < 2^{n-1}$, this is strictly positive, and so $P_{t}$ cannot be uniform.
\qed

\paragraph{Remark 3}
The condition $|\supp(\hat{f})| < 2^{n-1}$ in Theorem \ref{thm:connection} is tight.
This is because for $Q_{n}^{\eta} = \mathcal{B}_{n}(A_{f})$, 
where $\supp(f) = \{0_{n},\teta\}$ with $\eta = 1 \cdot \teta$, we have $\hat{f}(a) = 1 + \chi_{a}(\teta)$. 
This shows that $|\supp(\hat{f})| = 2^{n-1}$ and, by Theorem \ref{thm:eta}, $Q_{n}^{\eta}$ is uniform mixing.

\begin{corollary}
For $n \ge 2$, a continuous-time quantum walk on the join of two hypercubes 
$\mathcal{B}_{n}(J) = Q_{n} + Q_{n}$ is not instantaneous uniform mixing.
\end{corollary}
\prf The connection matrix $J$ is defined by the constant function $f \equiv 1$.
Thus, $\hat{f}(a) = \iverson{a = 0_{n}}$ and $|\supp(\hat{f})| = 1$.
By Theorem \ref{thm:connection}, $\mathcal{B}_{n}(J)$ is not uniform mixing.
\qed\\ 

\begin{figure}[h]
\begin{center}
\epsfig{file=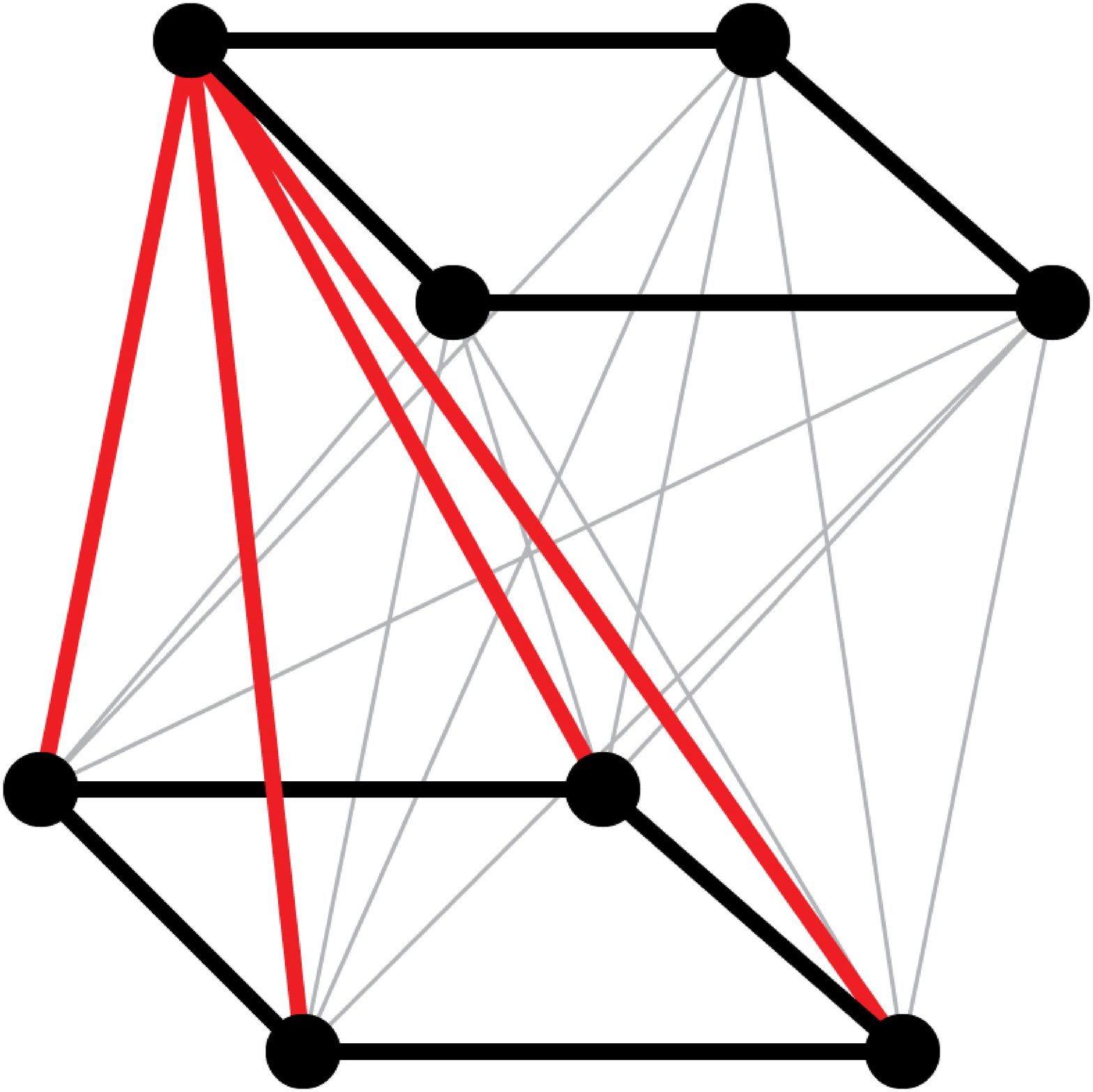, height=1.25in, width=1.25in}
\hspace{1in}
\epsfig{file=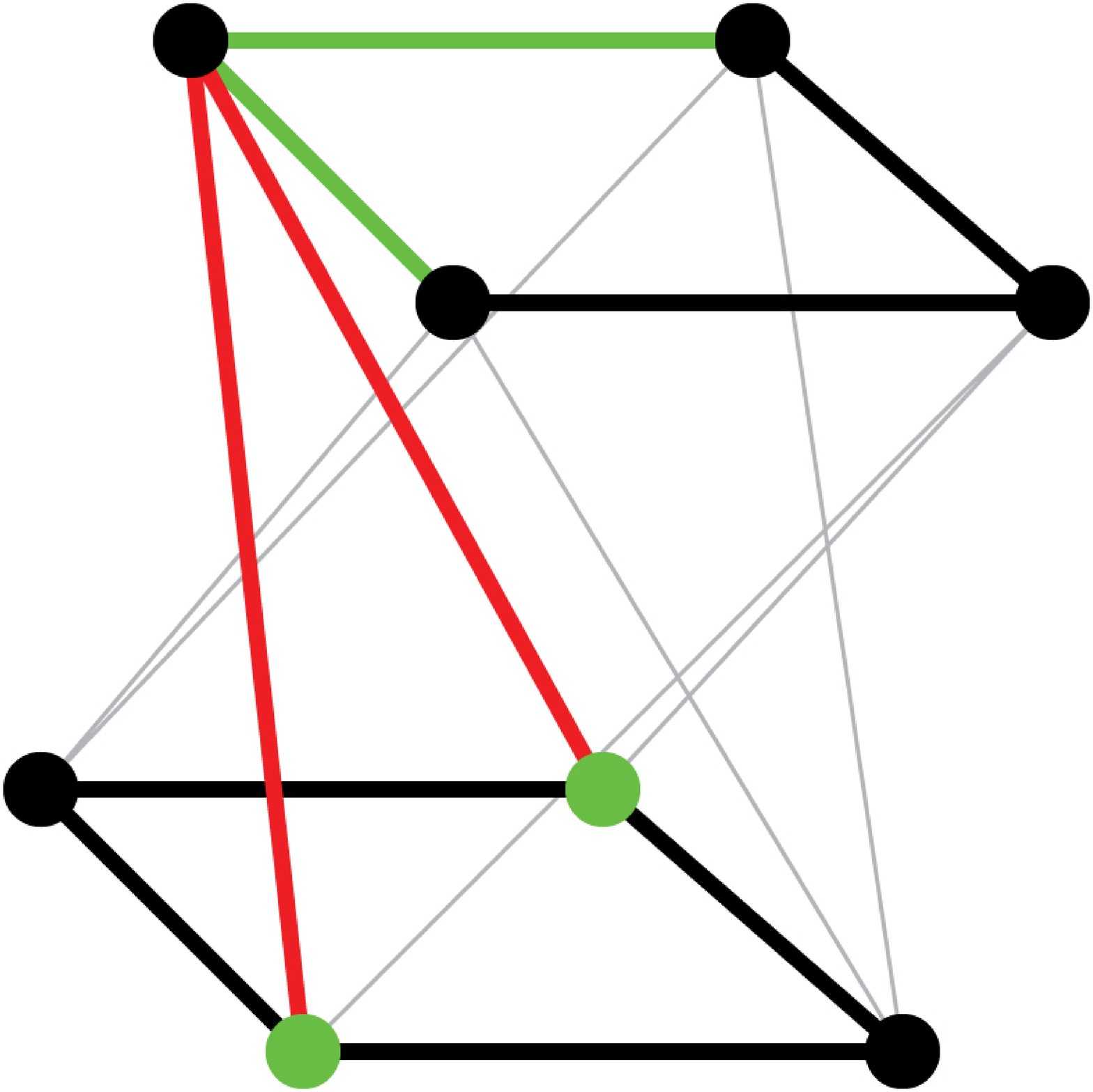, height=1.25in, width=1.25in}
\caption{Examples of hypercube bunkbeds: 
(a) $\mathcal{B}_{2}(J) = Q_{2} + Q_{2}$.
(b) $\mathcal{B}_{2}(Q_{2})$.
}
\label{figure:join-cube}
\end{center}
\end{figure}

\par\noindent
The next result shows that $|\supp(\hat{f})| < 2^{n-1}$ in Theorem \ref{thm:connection}
is not a necessary condition for $\mathcal{B}_{n}(A_{f})$ to be non-uniform mixing.

\begin{theorem}
For $n \ge 2$, a continuous-time quantum walk on $\mathcal{B}_{n}(Q_{n})$ is not instantaneous uniform mixing.
\end{theorem}
\prf
The adjacency matrix $A$ of $\mathcal{B}_{n}(Q_{n})$ is $J_{2} \otimes Q_{n}$.
The eigenvectors of $A$ are $\ket{\chi_{a}} = \ket{\chi_{a_1}} \otimes \ket{\chi_{\tilde{a}}}$, 
with corresponding eigenvalues 
$\lambda(a_{1} \cdot \tilde{a}) = (1+(-1)^{a_{1}})\lambda_{Q_n}(\tilde{a})$.
The amplitude of the quantum walk is given by
$\psi_{t}(a) = \exp\left( -it(1+(-1)^{a_{1}})\lambda_{Q_n}(\tilde{a})\right)$.
Using the Convolution Theorem, we obtain for $a = a_{1} \cdot \tilde{a}$ and
$b = b_{1} \cdot \tilde{b}$:
\begin{equation}
\widehat{P_{t}}(a) =
	\frac{1}{2^{n+1}} 
	\sum_{b} \exp\left[ -it \left\{(1+(-1)^{a_{1}+b_{1}})\lambda_{Q_n}(\tilde{a} \oplus \tilde{b}) 
	- (1+(-1)^{b_{1}})\lambda_{Q_n}(\tilde{b})\right\} \right]
\end{equation}
When $a_{1} = 0$, this yields
\begin{eqnarray}
\widehat{P_{t}}(a) 
	& = & \frac{1}{2} + 
		\frac{1}{2^{n+1}} \sum_{\tilde{b}} 
		\exp\left[ -2it\left(\lambda_{Q_n}(\tilde{a} \oplus \tilde{b}) 
		- \lambda_{Q_n}(\tilde{b})\right) \right] \\
	& = & \frac{1}{2} + \frac{1}{2} \cos\left( 2t \right)^{|\tilde{a}|}.
\end{eqnarray}
For even $|\tilde{a}|$, $\widehat{P_{t}}(a) \neq 0$.
Thus, $P_{t}$ never equals the uniform distribution.
\qed\\

\paragraph{Remark 4}
For $\mathcal{B}_{n}(Q_{n})$, we have $f(x) = \iverson{|x|=1}$ and hence $\hat{f}(a) = n-2|a|$.
Thus, $|\{a: \hat{f}(a)=0\}| = \binom{n}{n/2} \sim 2^{n}/\sqrt{n}$ which implies 
$|\supp(\hat{f})| \ge 2^{n-1}$. This shows that the condition $|\supp(\hat{f})| < 2^{n-1}$ in
Theorem \ref{thm:connection} is not a necessary condition.


\section{Conclusion}

In this paper, we have studied quantum uniform mixing on generalized hypercubes. 
Our first generalized hypercube is the bunkbed graph $\mathcal{B}_{n}(A_{f})$, where $A_{f}$ is
a $\zz^{n}$-circulant defined by a Boolean function $f: \zz^{n} \rightarrow \zo$. 
The second generalized hypercube we consider is the Hamming graph $H(n,q)$ or $q$-ary hypercube.
Following the work of Moore and Russell on the hypercube,
our main goal is to characterize which of these generalized hypercubes are uniform mixing and which are not. 

Prior to this work, there was only one known collection of graphs that is instantaneous uniform mixing. 
This is the class of Hamming graphs $H(n,q)$ with $q \in \{2,3,4\}$ (see \cite{mr02,cfhrtw07}). 
To this collection, we have added $\mathcal{B}_{n}(A_{f})$, where 
$\supp(f) = \{0_{n},a\}$ with $|a|$ odd.
This shows that $\mathcal{B}_{n}(A_{f})$ is uniform mixing if $0 < |\supp(f)| \le 2$ 
(with the aforementioned condition when $|\supp(f)|=2$);
thus generalizing the result of Moore and Russell \cite{mr02}.
We also showed that if  $|\supp(\hat{f})| < 2^{n-1}$, then $\mathcal{B}_{n}(A_{f})$ is not uniform mixing.
On the Hamming graphs, we have shown a tight characterization of quantum uniform mixing. 
Our main result states that $H(n,q)$ is not uniform mixing if and only if $q \ge 5$. 
This closes the gap left open from an earlier observation in \cite{cfhrtw07}.

Unlike the Hamming graphs, the bunkbed framework does not yield a tight characterization, 
since a counterexample is supplied by $\mathcal{B}_{n}(Q_{n})$. The Fourier techniques we employed 
do not seem powerful enough to prove a tight characterization based solely on the support sizes of $f$ 
and $\hat{f}$. Also, we suspect that quantum uniform mixing on the hypercube $Q_{n}$ is robust under 
the addition of up to $O(n)$ specific matchings. We leave these as open problems for future work.


\section*{Acknowledgments}

This research was supported in part by the National Science Foundation grant DMS-0646847
and also by the National Security Agency grant 42642.


\end{document}